# A novel site-controlled quantum dot system with high uniformity and narrow excitonic emission

L.O. Mereni, V. Dimastrodonato, R. J. Young and E. Pelucchi

*Tyndall National Institute, University College Cork, Cork, Ireland*

We report on the optical properties of a newly developed site-controlled InGaAs dots in GaAs barriers grown in pre-patterned pyramidal recesses by metalorganic vapour phase epitaxy. The inhomogeneous broadening of excitonic emission from an ensemble of quantum dots is found to be extremely narrow, with a standard deviation of 1.19 meV. A dramatic improvement in the spectral purity of emission lines from individual dots is also reported (18-30 µeV) when compared to the state-of-the-art for site controlled quantum dots.

**1. Introduction**

Studies on quantum dots (QDs) in connection to quantum optics have been constantly growing over the last ten years. The most challenging experiments have been performed until now *mainly* at the single dot level using randomly distributed QD ensembles, formed using the Stranski-Krastanow (SK) growth mode and, in most cases, using molecular beam epitaxy (MBE).[1] It is however important, in order to bring quantum dots closer to practical applications in the field of (optical) quantum communication and information processing, to build ensembles of QDs showing more controlled properties. Of particular significance, within the others, are 1) to obtain large ensemble of "identical" dots 2) the control on the absolute and relative position of quantum dots (which for example has to be controlled to enable the integration of quantum dots in a microcavity or in a photonic integrated circuit), 3) the deterministic matching of their emission wavelengths to the requested application and experimental environment, 4) to obtain dots with long or, ideally, no decoherence time.

We introduce here a *site-controlled* quantum dot system grown by metalorganic vapour phase epitaxy (MOVPE) in large (7.5 μm) inverted tetrahedral recesses on (111)B GaAs substrates. This represents an evolution to the InGaAs pyramidal QD system which raised considerable interest in the community for its versatility[2] and unifomity.[3] We show that it is possible to replace the $Al_yGa_{1-y}As$ barrier material with GaAs. This delivers improved epitaxial quality and, critically, preserves the delicate equilibrium between growth rate anisotropy and capillarity processes that leads to a self limiting profile and, finally, the dot formation.[4] We present the results of optical studies of this system, in particular we show: the ability to tune the emission wavelength, a neutral exciton emission of record spectral purity for quantum dots grown by MOVPE and/or "site controlled" and a record high uniformity of the emission energy of the neutral exciton.

## 2. Results and discussion

Our samples were grown by MOVPE at low pressure (20 mbar) in a commercial horizontal reactor. Standard precursors, trimethyl-Aluminim/Gallium/Indium, and purified $AsH_3$, were used in purified $N_2$ carrier gas. The epitaxial growth of the layers occurs on GaAs (111)B ±0.1° substrate which is *pre-patterned*, using standard UV lithography and wet chemical etching, with an array of 7.5μm pitch inverted tetrahedral pyramids. As result of the selective etching process, the pyramids expose sharp lateral facets with crystallographic orientation (111)A. Growth conditions, temperature (730 ºC thermocouple), V/III ratio (in the range 600-800), and the material structure surrounding the quantum dot layer were nominally the same for all samples. Following a buffer layer of GaAs, and $Al_yGa_{1-y}As$ layers which act as an etch stop to aid the post-growth process of substrate removal, the quantum dot and barrier layers are grown sandwiched between $Al_{0.55}Ga_{0.45}As$ cladding to provide carrier confinement. The width of the GaAs barrier beneath (above) the dot layer was 100 (70) nm. Four samples (A-D) were grown for this study each with a different composition of $In_xGa_{1-x}As$ in their respective 0.5 nm thick quantum dot layers; for samples A, B, C and D the compositions respectively were, x = 0.15, 0.25, 0.35 and 0.45.[5] We underline that, while comprising a similar material system, this design is physically different to the one reported recently in Ref. 6. In the work by Gallo et al. the dimensions of the pyramidal template are >20 times smaller than in ours and, critically, of the order of the adatom diffusion length, thus the growth dynamics in their system and results they find are quite different to those presented here.

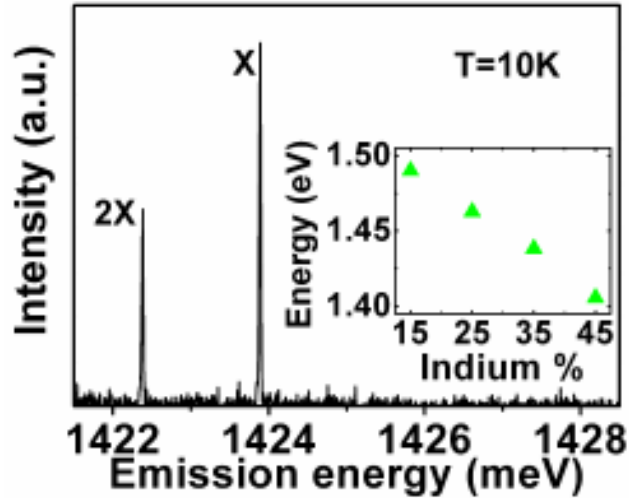

*Figure 1*

Figure 1 (main) shows the exciton (X) and biexciton (2X) emission spectrum (peaks assignment was made by studying a power dependent peaks dynamic) from our best pyramidal $In_{0.25}Ga_{0.75}As$ QD, whose nominal thickness is 0.5nm, measured in the apex-up geometry after the typical, for these kinds of pyramidal QDs systems, "back-etching" substrate removal was performed.[4,7] The FWHM of the free exciton peak was as narrow as 18µeV, demonstrating excellent optical properties and low spectral meandering (which also is an indication, to some extent, of reduced decoherence probabilities). It is important to underline that most of the QDs on the sample showed anyway very narrow transition lines, mostly below 30 µeV. These results represent a record for all site controlled QDs [8] and are comparable with those obtained from high quality MBE grown SK QDs.

Figure 1 (inset) shows the energy shift as function of In content in the dot layer: for samples 1-4, with a nominal dot layer thickness of 0.5nm. A total energy shift of ~100 meV

L. Mereni et al.

was obtained. This relatively low value suggests that the thickness of our QDs is small enough that the exciton wavefunction is only partially confined in the dot layer. Future theoretical studies will help clarifying this point.

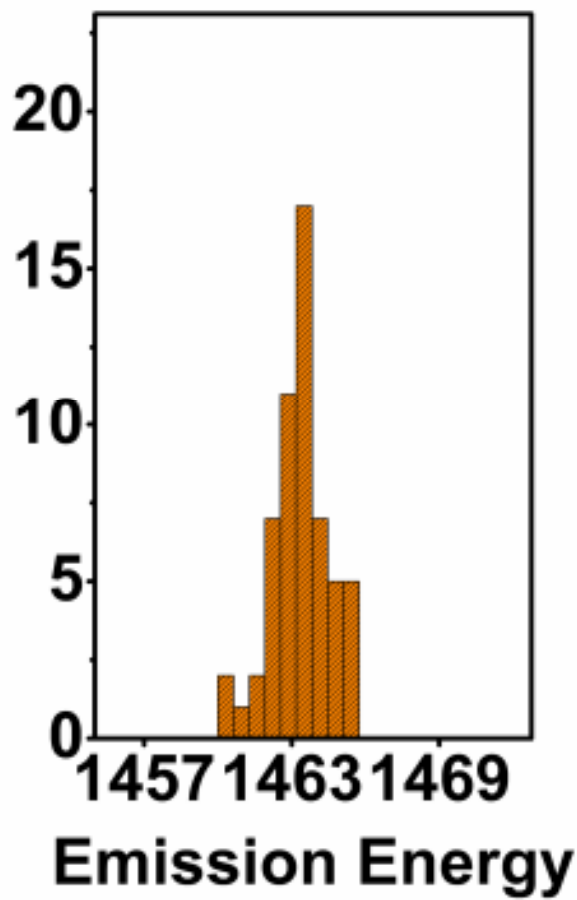

*Figure 2*

The uniformity in the emission wavelength from sample B, measured in the as-grown geometry, is illustrated in Fig. 2 which plots the neutral exciton energy distribution for *a large number* (~ 60) of individual pyramids. The average emission energy of the neutral exciton line, was found to be 1463.5 meV, while the FWHM of the corresponding

L. Mereni et al.

interpolated Gaussian fit is 2.8 meV. To better appreciate the importance of this result it must be observed that, in general, to compare the different dot ensembles reported in the literature, it is useful to consider the inhomogeneous broadening of an ensemble of QDs as a function of the confinement energy, or a related parameter. In fact, smaller QDs typically exhibit larger variations in their transition energies for a given size variation if compared to bigger dots. Figure 3 plots the inhomogeneous broadening of a collection of different sets, from the literature, of (In)GaAs/(Al)GaAs SK and Pyramidal QDs, obtained from luminescence experiments, against the fundamental to first excited state/barrier level separation (S-P), chosen here as a "measure" of the confinement energy.[3,9,10,11,12] As discussed, the inhomogeneous broadening generally increases with increasing confinement. Rapid thermal annealed SK dots can show extremely high uniformities; however Pyramidal QDs tend to show much smaller size broadening for a given S-P separation, and should probably be considered as the most uniform QD system to date. An inhomogeneous broadening as low as ~ 4meV was achieved with S-P separation as large as ~ 30meV in the group of Prof. Kapon in Lausanne, while our best sample (filled triangle) shows an even higher uniformity while preserving (~ 40 meV) a high confinement.

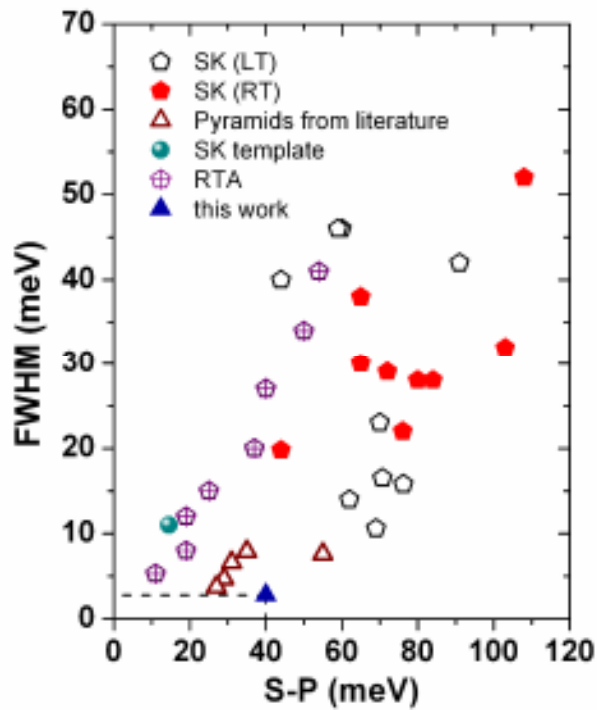

*Figure 3*

In conclusion we have introduced a site-controlled quantum dot system, composed of an $In_xGa_{1-x}As$ dot embedded in GaAs barriers, demonstrating both high uniformity and spectral purity. Simply varying the composition of the $In_xGa_{1-x}As$ quantum dot allowed the emission energy of the neutral exciton to be tuned and controlled. We believe these results are an encouraging sign that this new kind of Pyramidal quantum dot will be very useful for fundamental quantum optics and physical studies.

**Acknowledgments**

Science Foundation Ireland under grants 05/IN.1/I25 and 05/IN.1/I25/EC07. We are grateful to K. Thomas for his support with the MOVPE system.

---

X. Zhao, and M. Sadeghi, Appl. Phys. Lett. 81, (2002) 1621; B. Alloing and A. Fiore et al., private communication.

[12] A. Rastelli, S. Stufler, A. Schliwa, R. Songmuang, C. Manzano, G. Costantini, K. Kern, A. Zrenner, D. Bimberg, and O. G. Schmidt, Phys. Rev. Lett. 92, (2004) 166104.

L. Mereni et al.

## Figure captions

**Figure 1** (*colour online*) -*main:* µ-Photoluminescence spectra taken from two different individual quantum dots. Emission from the neutral exciton (X) and biexciton (2X) states is labelled. The extremely low value of the linewidth of the excitonic emissions was extracted by fitting a lorentzian peak. Integration times and excitation intensities are different for the two spectra. –*Inset:* Emission energy tuning as a function of the Indium concentration in the dot layer.

**Figure 2** (*colour online*) Uniformity of the emission of the neutral exciton line on the sample with 25% Indium content in the alloy of the dot layer. The standard deviation of the Gaussian distribution is only 1.2 meV corresponding to a FWHM of 2.8 meV.

**Figure 3** (*colour online*) Selection of literature data on the FWHM of luminescence spectra versus the S-P transitions energy separation for SK and pyramidal QDs. SK(LT): low temperature luminescence data for SK dots [Ref. 9, 10]; SK(RT): room temperature luminescence data for SK dots [Ref.11]; Pyramids from literature: low temperature PL and cathodoluminescence data for pyramidal QDs [Ref. 3]; SK template: low temperature PL data for GaAs QDs grown inside etched SK dot template [Ref. 12]; RTA: low temperature PL data for SK dots after rapid thermal annealing [Ref. 10]; this work: our best sample.